\documentclass[pra,twocolumn,superscriptaddress,nofootinbib]{revtex4-2}
\usepackage{graphicx}
\usepackage{bm}
\usepackage{amssymb,amsmath,mathrsfs}
\usepackage{color}
\usepackage{amstext}
\usepackage[english]{babel}
\usepackage{latexsym}
\usepackage{array}   
\usepackage{multirow}         
\usepackage[usenames,dvipsnames]{xcolor}
\usepackage[colorlinks=true,citecolor=Blue,linkcolor=RubineRed,urlcolor=Blue]{hyperref}
\usepackage[all]{xy}     
\usepackage[usenames,dvipsnames]{xcolor}

\definecolor{mygreen}{RGB}{0,110,62}
\definecolor{mypurple}{RGB}{134,3,191}

\begin{document}

\title{Supersymmetry in nonlinear and linear Quantum Optics: the Kerr-like and multiphoton Jaynes-Cummings models}

\author{Ivan~A. Bocanegra-Garay\href{https://orcid.org/0000-0002-5401-7778}{\includegraphics[scale=0.45]{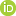}}}
\email[]{ivanalejandro.bocanegra@uva.es}
\affiliation{Departamento de F\'isica Te\'orica, At\'omica y \'Optica and  Laboratory for Disruptive Interdisciplinary Science (LaDIS), Universidad de Valladolid, 47011 Valladolid, Spain}

\author{L. Hernández-Sánchez\href{https://orcid.org/0009-0008-2648-4353}{\includegraphics[scale=0.45]{orcid}}}
\affiliation{Instituto Nacional de Astrof\'isica \'Optica y Electr\'onica, Calle Luis Enrique Erro No. 1\\ Santa Mar\'ia Tonantzintla, Puebla, 72840, Mexico}

\author{H. M. Moya-Cessa\href{https://orcid.org/0000-0003-1444-0261}{\includegraphics[scale=0.45]{orcid}}}
\affiliation{Instituto Nacional de Astrof\'isica \'Optica y Electr\'onica, Calle Luis Enrique Erro No. 1\\ Santa Mar\'ia Tonantzintla, Puebla, 72840, Mexico}

\date{\today}

\begin{abstract}
A novel approach is proposed to analyze a rather vast counter-rotating Hamiltonian interaction in the context of cavity quantum electrodynamics. The method relies upon the supersymmetric mapping of the corresponding rotating interaction and allows the analysis of the dynamics in the counter-rotating system in a fully general and exact analytical manner. Intriguing features of the counter-rotating system are revealed through the simple supersymmetric transformation. In turn, such interesting attributes have an important range of potential technological applications. In this way, supersymmetry emerges as a useful tool to both connect and construct exactly solvable photonic systems in cavity quantum electrodynamics, and more generally in quantum optics, as well as to analyze the corresponding physical consequences and their possible technology implementations.
\end{abstract}
\maketitle

\section{Introduction}
The interaction between light and matter has been an important topic of study from both theoretical and experimental points of view in the last decades \cite{JC_Model63,Cummings_65,Eberly_80,Shore_93,Larson_2021,Brune_PRL96}. Indeed, cavity quantum electrodynamics (cQED) is currently an active and prolific field of research, whose main applications cover trendy topics such as quantum information, quantum computing, material engineering, quantum control, among others \cite{Diedrich_PRL89,Blockley_ELet92,Wallraff_Nat04,Solano_PRL03,Lara_PRA05,Bermudez_PRA07,Goldman2009,Mischuck_PRA13,Liu2021,Ritsch_RModPhys13,Blais_RModPhys21}. In that context, nonlinear mechanisms have been object of intense research: Kerr nonlinearities and field-intensity dependent couplings are a pair of cases of interest \cite{Buck_81,Buzek_90,Moya_95Kerr,Cordero_2011,Kitagawa_86,Tara_93,Werner_91,Rivera_97,Chumakov_99}. In parallel, multiphotonic processes (where the qubit transitions occur through the absorption or emission of more than one photon \cite{Sukumar_81,Gerry_88,Joshi_92,Deppe_2008,Gonzalez_2013,Villas_2019,Zou_2020}, see Figure \ref{fig_1}) have also played an important role, exhibiting greater richness and involved dynamics,  compared with the single-photon interactions in the Jaynes-Cummings (JC) model, for instance.

Certainly, in the framework of cQED it is well known that the rotating-wave approximation (RWA) allows dropping the \textit{counter-rotating terms} (see Figures \ref{fig_1}E and \ref{fig_1}F, for the case of multiphotonic interaction). In the case of the quantum Rabi model, the RWA leads to the simpler (integrable) JC model involving only rotating terms (Figures \ref{fig_1}C and \ref{fig_1}D, when $k=1$). Although the counter-rotating interaction has been referred to as the non-conservation of excitations (see for instance the textbook of Gerry and Knight \cite{Gerry2004}), such effects are indeed being produced in actual laboratory environments and are significant in valuable contexts as ion-trapping and the production of entangled states \cite{Solano_PRL03,Lara_PRA05, Bermudez_PRA07,Goldman2009,Mischuck_PRA13,Liu2021}. In turn, these are relevant in an ample extension of areas ranging from basic science to applied physics: quantum computing and quantum information, for example.

In a different context, supersymmetric quantum mechanics (SUSY QM) has allowed to connect physical models utilizing supersymmetric (SUSY) operators, namely as mappings between the corresponding Hilbert spaces \cite{Cooper1995,fernandez1999,Mielnik2000,Samsonov_2004,Hussin_2006,miri012013,David_14,Correa2015,Mig032019,BocanegraGaray2024}. Indeed, SUSY transformations are an efficient tool to construct exactly solvable Hamiltonians in time-independent as well as time-dependent scenarios. In this manner, supersymmetry serves as a profitable framework for both linking and designing exactly solvable photonic systems in cavity quantum electrodynamics, as well as for testing the associated physical implications and their prospective technological uses.

Specifically, this work focuses on the case of a SUSY operator that connects Hamiltonians separately involving rotating and counter-rotating terms in a general nonlinear and multiphotonic interaction, thus filling a significant void in the available literature. Two particular cases are of special interest: the nonlinear Kerr interaction and the multiphoton (MP) Jaynes-Cummings model introduced by Sukumar and Buck in 1981. It is worth emphasizing that SUSY QM has just begun to be explored in the context of cQED and quantum optics \cite{Maldonado2021,Kafuri2024,Bocanegra_2024PRR}. Consequently, the SUSY connection developed in this study represents a noteworthy contribution that should not be underestimated.

The contents of the work are given as follows. Section \ref{sec_1} introduces the model considering a general rotating interaction between a cavity field and a single qubit (the extension to N qubits is direct). In Section \ref{sec_2}, the SUSY formalism is reviewed and seen to generate the counter-rotating interaction we are interested in. Besides, it is shown that the SUSY transformation defines a mapping from the solutions of the rotating system into those of the counter-rotating one. Thus, Section \ref{sec_2} is certainly the core of the current manuscript. In Section \ref{sec_3}, exact analytical expressions for dynamical variables of interest of the counter-rotating system are supplied. In turn, in Section \ref{sec_4} some numerical examples are presented, illustrating intriguing features of the counter-rotating system, such as the production of quantum (sub-Poissonian) states of light as a consequence of the counter-rotating MP interaction, as well as the creation of generalized Schr\"odinger cat states and the periodic recovery of the initial cavity-field state in the nonlinear Kerr scenario. In Section \ref{sec_5}, the main conclusions are drawn and perspectives for future work are sketched out. For completeness, a set of Appendices is also provided.

\section{Initial model: the rotating interaction}\label{sec_1}
\begin{figure}[t!]\
\centering
\includegraphics[width=1\linewidth]{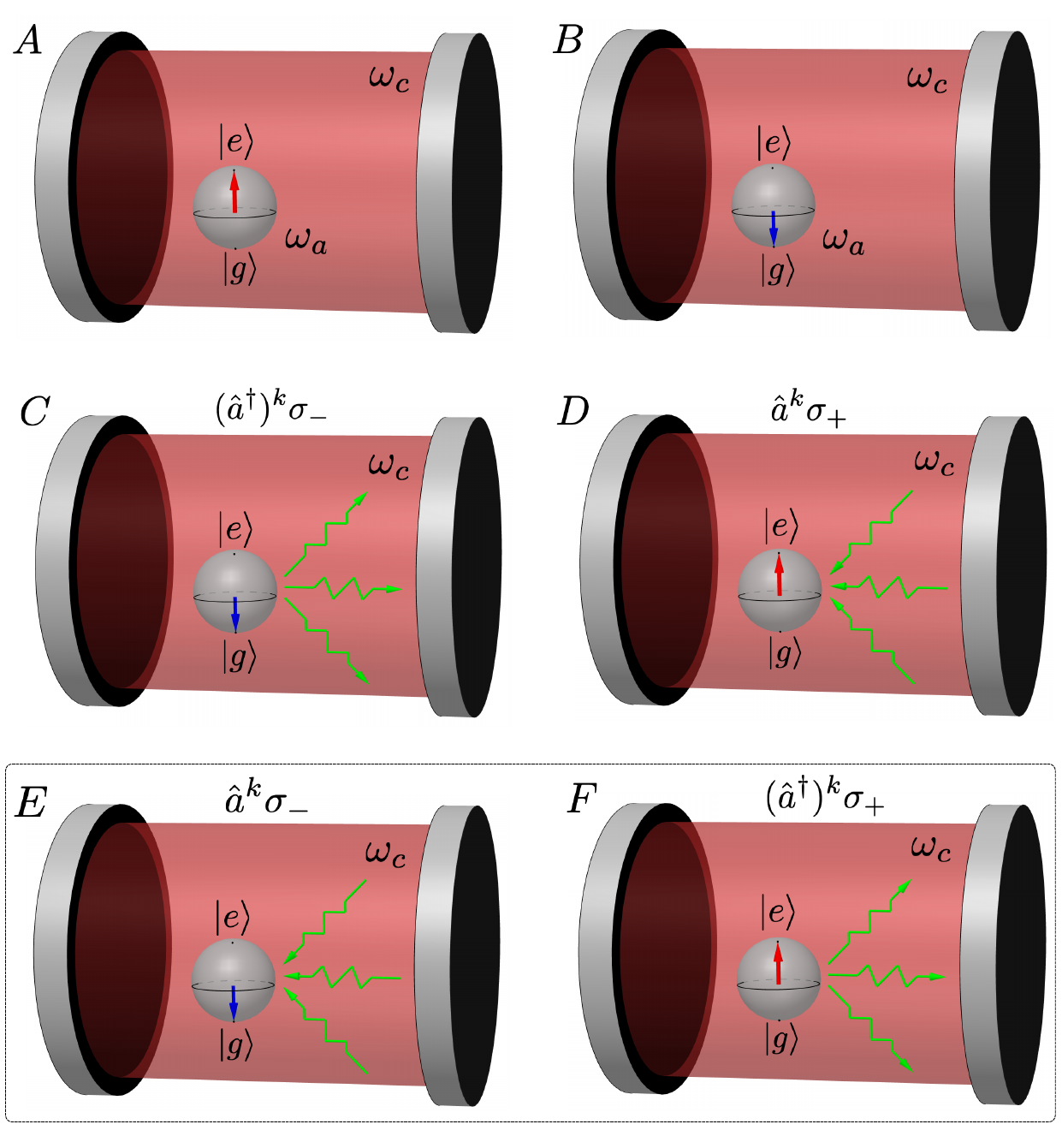}
\caption{\label{fig_1}  {\bf Schematic representation of the effect of rotating and counter-rotating terms in cQED.} A qubit in its excited (ground) state is embedded in a high-Q cavity, as shown in A (B). The effect of the rotating terms on the state in A (B) is exemplified in C (D): the transition of the qubit to its ground (excited) state leads to the creation (annihilation) of $k$ photons (green arrows). On the other hand, the action of the counter-rotating terms
on the initial state A (B) is represented in E (F): the transition of the qubit to its ground (excited) state is accompanied by the destruction (creation) of $k$ photons. These last processes have been referred to as \textit{non-conservative}, although they appear in actual physical systems. In turn, the transition between the rotating and counter-rotating interactions (middle and lower rows, respectively) is straightforward by means of the SUSY technique, as explained in the main text.} 
\end{figure}

A general model considering a rather broad type of interaction between a two-level system (a qubit) embedded in a high-Q (lossless) cavity is taken into account (as shown in the upper row of Figure \ref{fig_1}). The Hamiltonian in the interaction picture (using $\hbar =1$) is expressed as 
\begin{equation}\label{gen_H}
	\hat H_0 = \left(\frac{\Delta}{2} + F_{\hat n}\right)\hat\sigma_z + G_{\hat n} + g[\hat a^k f_{\hat n} \hat\sigma_+ + f_{\hat n} (\hat a^\dagger)^k \hat\sigma_-],
\end{equation}
with $k = 1,2,\dots$, and where $\Delta$ is the detunning between the cavity and atomic frequencies, $\omega_c$ and $\omega_a$, respectively, $F_{\hat n}, G_{\hat n}$ and $f_{\hat n}$ are arbitrary functions of the number operator $\hat n = \hat a^\dagger \hat a$, where $\left\{\hat a^\dagger, \hat a\right\}$ are the usual bosonic operators $[\hat a,\hat a^\dagger]=\mathbb{I}$, $g$ is the qubit-cavity coupling whenever $f_{\hat n}= \mathbb I$, and $\left\{\hat\sigma_z,\hat\sigma_\pm\right\}$ are pseudo-spin operators satisfying $[\hat{\sigma}_{+}, \hat{\sigma}_{-}] = \hat{\sigma}_{z}$ and $[\hat{\sigma}_{z}, \hat{\sigma}_{\pm}] = \pm 2 \hat{\sigma}_{\pm}$. The qubit operators can be expressed in the base of the excited $|e\rangle$ and ground $|g\rangle$ states of the atom as $\hat{\sigma}_z=\vert e\rangle\langle e\vert -\vert g\rangle\langle g\vert$, $\hat{\sigma}_{-}=\vert g\rangle\langle e\vert$, and $\hat{\sigma}_{+}=\vert e\rangle\langle g\vert$. There is a constant of motion $\hat C_0 = \hat a^\dagger \hat a + k\frac{\hat\sigma_z}{2}$
associated with $\hat H_0$ in (\ref{gen_H}), i.e. $[\hat H_0,\hat C_0] = 0$.

In general terms, $k$ allows for multiphoton interaction, and the function $f_{\hat n}$ represents a coupling between the cavity and the qubit depending on the intensity of the cavity field. Besides, a nontrivial $F_{\hat n}$ induces a generalized Stark shift interaction, as ($F_{\hat n} = \hat n$) the one caused by non-resonant levels of the atomic state \cite{Moya_91,Moya_95Kerr}. Finally, $G_{\hat n}$ can be chosen to model Kerr nonlinearities ($G_{\hat n} = \hat n^2$), parity effects ($G_{\hat n} =(-1)^{\hat n}$), etc. (see for instance Ref. \cite{Dehghani_2016}).

As mentioned in the introductory section, this paper focuses on two particular, yet relevant, cases. The nonlinear Kerr-like Hamiltonian ($F_{\hat n}=0, G_{\hat n}=\chi \hat n^2$, $f_{\hat n}=\mathbb I$ and $k=1$) reads 
\begin{equation}\label{KRR}
 \hat{H}_{\rm Kerr} 
 = \frac{\Delta}{2}\hat{\sigma}_z +\chi \hat n^2 +g[\hat{a}\hat{\sigma}_{+}+\hat{a}^{\dagger}\hat{\sigma}_{-}].
\end{equation}
In turn, the multiphoton Jaynes-Cummings Hamiltonian ($F_{\hat n}=G_{\hat n}=0, f_{\hat n} = \mathbb I$, $k\geq 1$) is given by
\begin{equation}\label{PA_JC}
 \hat{H}_{\rm MP} 
 = \frac{\Delta}{2}\hat{\sigma}_z+g[\hat{a}^k\hat{\sigma}_{+}+(\hat{a}^{\dagger})^k \hat{\sigma}_{-}].
\end{equation}
Note that the previous Hamiltonians only consider the rotating terms (see also the middle row of Figure \ref{fig_1}). After a SUSY transformation, it is possible to turn such Hamiltonians into those considering the counter-rotating terms instead (lower row of Figure \ref{fig_1}) and to map the corresponding solutions. In the following section, the SUSY formalism is presented, showing that is feasible to define a mapping between the Hilbert spaces associated with the rotating and counter-rotating systems in a fairly straightforward fashion.

\section{Supersymmetry and the counter-rotating interaction}\label{sec_2}

From the SUSY QM point of view, provided the solution $|\psi(t)\rangle$ of the Schr\"odinger equation $i\partial_t |\psi(t)\rangle = \hat H_0 |\psi(t)\rangle$, for a given time-independent Hamiltonian $\hat H_0$, if there exists an (\textit{intertwining} or SUSY) operator $\hat{\mathcal A}$ satisfying the relation
\begin{equation}\label{susy_t}
  \hat{\mathcal A}\hat H_0 = \hat H\hat{\mathcal A},  
\end{equation}
then the solution $|\phi(t)\rangle\propto\hat{\mathcal A}|\psi(t)\rangle$ of the Schr\"odinger equation $i\partial_t |\phi(t)\rangle = \hat H |\phi(t)\rangle$ is determined as well.

For $\hat H_0$ as given in (\ref{gen_H}), and $\hat{\mathcal A} = \mathcal{\hat B}^k$, where
\begin{equation}\label{bk}
    \mathcal{\hat B}^k = \hat\sigma_+\hat\sigma_- (\hat a^\dagger)^k + \hat\sigma_-\sigma_+ \hat a^k,
\end{equation}
the Hamiltonian 
\begin{equation}\label{SUSYp}
    \hat H = \frac{\Delta}{2}\hat\sigma_z + \hat S_+ + \hat S_- + g[\hat a^k f_{\hat n} \hat\sigma_- + f_{\hat n} (\hat a^\dagger)^k\hat\sigma_+],
\end{equation}
satisfying (\ref{susy_t}), is called a SUSY \textit{partner} of $\hat H_0$. The operators $\hat S_\pm \equiv \hat\sigma_\pm \hat\sigma_\mp(G_{\hat n \mp k\mathbb I} \pm F_{\hat n \mp k\mathbb I})$ were defined. Thus, the known solutions associated with (\ref{gen_H}) can be straightforwardly mapped into those corresponding to (\ref{SUSYp}), through the action of the operator (\ref{bk}). The constant of motion $\hat C$ associated with Hamiltonian (\ref{SUSYp}), this is fulfilling $[\hat H,\hat C] = 0$, is simply $\hat C = \hat a^\dagger\hat a -k\frac{\hat\sigma_z}{2}$. In Appendix \ref{appendix_matrix}, a matrix representation of the previous operators in the qubit base is given, from which is easy to see that $\hat H_0, \hat H$ and $\mathcal{\hat B}^k$, satisfy the intertwining relation (\ref{susy_t}). As noted before, nontrivial supersymmetry has only started to be explored in the context of cQED. Therefore, the set $\left\{\hat H_0, \hat H, \mathcal{\hat B}^k\right\}$ should not be undervalued.

Indeed, the SUSY transformation (\ref{susy_t}) naturally produces the emergence of counter-rotating terms in the interaction part of the Hamiltonian (\ref{SUSYp}). Then, supersymmetry arises as an appealing tool for studying the effects of the counter-rotating interaction in the framework of the general qubit-atom interaction described by (\ref{SUSYp}). In turn, the interplay between rotating and counter-rotating interactions is achievable in actual laboratory schemes, employing frequency shifts in the interacting laser \cite{Solano_PRL03,Lara_PRA05, Bermudez_PRA07}. Also mentioned, such counter-rotating interaction becomes important in the context of ion-trapping and the production of non-classical entangled states. In addition, the counter-rotating terms can be awarded for the presence of external sources of electromagnetic field; in this context, the counter-rotating systems seem as good candidates to model open quantum systems as well \cite{Solano_PRL03}. Thus, the analysis of Hamiltonians involving counter-rotating interactions is something worthwhile by itself, with promising applications in several fields of nowadays physics.

Consequently, the potential of SUSY QM as a convenient apparatus to both link and design exactly solvable photonic quantum systems in cQED as well as for evaluating the associated physical implications together with their possible technological implementations is something undeniable and deserves attention. In the next section, some specific physical quantities of interest in the counter-rotating system are analyzed, by taking advantage of the SUSY connection between the rotating and counter-rotating systems, and from the solution of the Schr\"odinger equation associated with the former and provided in Appendix \ref{appendix_solutions}.\\

\section{Dynamical variables of the counter-rotating system}\label{sec_3}
We are interested in the particularly simple, yet truly interesting, counter-rotating system defined by 
\begin{equation}\label{multi_kerr}
    \hat H = \frac{\Delta}{2} \hat\sigma_z + \chi \hat n^2 + g[\hat a^k \hat\sigma_- + (\hat a^\dagger)^k\hat\sigma_+],
\end{equation}
encompassing a nonlinear Kerr medium ($\chi\neq 0$) and multiphotonic ($k\geq 1$) scenarios. The corresponding evolution operator $\hat U(t) = \exp{(-i\hat H t)}$ is
\begin{equation}\label{evop}
    \hat{U} = \left(
\begin{matrix} 
\hat U_{11} & \hat U_{12}
\\ 
\hat U_{21} & \hat U_{22}
\end{matrix}
\right),
\end{equation}
with
\begin{equation}
\begin{aligned}
    \hat{U}_{11} &= \mathcal{E}_{\hat{n} - k\mathbb{I}} \mathcal{F}_{\hat{n} - k\mathbb{I}}^\dagger, 
    &\quad \hat{U}_{12} &= (\hat{a}^\dagger)^k \mathcal{E}_{\hat{n}} \mathcal{G}_{\hat{n}}, \\
    \hat{U}_{21} &= \hat{a}^k \mathcal{E}_{\hat{n} - k\mathbb{I}} \mathcal{G}_{\hat{n} - k\mathbb{I}}, 
    &\quad \hat{U}_{22} &= \mathcal{E}_{\hat{n}} \mathcal{F}_{\hat{n}},
\end{aligned}
\end{equation}
where the functions $\mathcal{E}_{\hat n}$, $\mathcal{F}_{\hat n}$ and $\mathcal{G}_{\hat n}$ are defined in Appendix \ref{appendix_solutions}. In turn, the evolution operator (\ref{evop}) was obtained using the relation $\hat{\mathcal A}\mathcal{\hat U} = \hat U\hat{\mathcal A}, $ with $\mathcal{\hat U}$ the evolution operator corresponding to the initial Hamiltonian $\hat H_0$ given in (\ref{Kerrdisp}). Therefore, the Schr\"odinger equation for the Hamiltonian (\ref{multi_kerr}) can be solved straightforwardly by employing the supersymmetric technique presented in the previous section. The general non-separable and normalized initial state (see Appendix \ref{appendix_entanglement})
\begin{equation}\label{non-separable}
        |\phi(0)\rangle = N_{\mathrm{eg}}^{-1}[\alpha_e|e\rangle\otimes\sum_n c_n|n\rangle
    +  \alpha_g|g\rangle\otimes\sum_n d_n|n\rangle],
\end{equation}
is considered and used to compute some of the dynamical variables of the counter-rotating system described by (\ref{multi_kerr}). In what follows, exact analytical expressions are given for the atomic inversion, defined as the expectation value of the $\hat\sigma_z$ operator; the expectation value of the $j$-th power ($j = 1, 2, \dots$) of the number operator $\hat n$, which contains the information of the mean number of photons in the cavity field; the Mandel $Q$ parameter, providing information about the classical or quantum nature of the cavity field; as well as the Wigner function and fidelity of the cavity field.

\subsection{Expectation values of diagonal operators and Mandel $Q$ parameter}

The expectation value of a diagonal operator 
\begin{equation}
    \mathcal{\hat O}
    = \begin{pmatrix}
    p_{\hat n} & 0\\
    0 & q_{\hat n}
    \end{pmatrix},
\end{equation}
namely $\langle \mathcal{\hat O} \rangle (t) = \langle\phi(t)| \mathcal{\hat O} |\phi(t)\rangle$, with $|\phi(t)\rangle = \hat U |\phi(0)\rangle$, where $p_{\hat n}, q_{\hat n}$ are arbitrary functions of the number operator, is straightforwardly obtained from the evolution operator (\ref{evop}) and the initial condition (\ref{non-separable}), as
\begin{equation}\label{expec_diagonal}
        \langle \mathcal{\hat O}\rangle = |N_{\mathrm{eg}}|^{-2}\sum_{n=0}^\infty\left\{ |\alpha_e|^2 \mathcal Q_n+|\alpha_g|^2 \mathcal R_n  + 2 \mathcal T_n\right\},
\end{equation}
with the following definitions
\begin{equation}
    \begin{split}
    \mathcal Q_n &= p_n |c_n|^2 |\mathcal F_{n-k}|^2 + q_n |c_{n+k}|^2 |\mathcal H_n|^2, \\
    \mathcal R_n &= q_n |d_n|^2| \mathcal F_n|^2 + p_{n+k} |d_n|^2 |\mathcal H_n|^2,\\
    \mathcal T_n &= \mathcal H_n(p_{n+k}-q_n)\mathrm{Im}[\alpha_e\bar\alpha_g c_{n+k}\bar d_n\bar{\mathcal F}_n], \\
    \mathcal H_n &= -i\sqrt{\frac{(n + k)!}{n!}} \mathcal G_n,
    \end{split}
\end{equation}
the bar denoting complex conjugation, $\mathcal F_n = \langle n|\mathcal F_{\hat n}|n\rangle$, and similarly for $\mathcal G_n$, $p_n$ and $q_n$. Note that $\Omega_m = \frac{\Delta}{2} + \chi(mk + \frac{k^2}{2})$ for negative $m$. Although very simple, the expression (\ref{expec_diagonal}) is useful in cases of practical interest (see Table \ref{table1}), for instance for obtaining the atomic inversion $\langle \hat\sigma_z \rangle(t)$, defined as the difference of the probabilities of the qubit being in the excited or the ground state, i.e. $\langle \hat\sigma_z \rangle = P_e - P_g$. It is also useful to compute the expectation value of the $j$-th power of the number operator $\langle \hat n^j \rangle (t)$. Besides, the Mandel $Q$ parameter, defined as \cite{Mandel_79}
\begin{equation}
    Q = \frac{\langle \hat n^2\rangle - \langle \hat n \rangle^2}{\langle \hat n \rangle} - 1,
\end{equation}
can be easily obtained from Table \ref{table1} and the expression (\ref{expec_diagonal}). The Mandel $Q$ parameter gives information about the Poissonian ($Q = 0$), sub-Poissonian  ($Q<0$), or super-Poissonian ($Q>0$) nature of the cavity field. In particular, a sub-Poissonian field is quantum, showing properties that have no classical analog.
\begin{table}[t!]
\centering
\caption{\label{table1} \bf{Diagonal operators in the qubit base.}}
\begin{tabular}{ >{\centering\arraybackslash}p{1in} >{\centering\arraybackslash}p{1in} >{\centering\arraybackslash}p{1in} }
\hline\hline
$\mathcal{\hat O}$ & $p_{\hat{n}}$ & $q_{\hat{n}}$ \\
\hline
$\hat{\sigma}_z$ & $\mathbb I$ & $-\mathbb I$ \\
$\hat n^j$ & $\hat n^j$ & $\hat n^j$ \\ 
\hline\hline
\end{tabular}
\end{table}

Finally, remark that (\ref{expec_diagonal}) describes the dynamics of a system that is initially in the non-separable state (\ref{non-separable}). However, in the case of an initially separable state, for instance, $c_n = d_n$ for all $n$, or $\alpha_g = 0$, $\alpha_e \neq 0$ ($\alpha_g \neq 0$, $\alpha_e = 0$), the expression (\ref{expec_diagonal}) becomes even simpler.

\subsection{Wigner function}
In addition, the Wigner function for the cavity-field reduced density matrix $\hat \rho_F(t) = \mathrm{Tr}_A [\hat\rho(t)]$, with $\hat\rho(t) = |\phi(t)\rangle \langle \phi(t)|$, is defined as \cite{Moya_book}
\begin{equation}
    W (\alpha,\alpha^*)= \frac{1}{\pi}\sum_{s=0}^\infty (-1)^s\langle s|\hat D_\alpha^\dagger \hat\rho_F(t)\hat D_\alpha |s\rangle,
\end{equation}
where $\hat D_\alpha = \exp(\alpha \hat a^\dagger - \bar\alpha\hat a)$ is the Glauber displacement operator. The cavity-field reduced density matrix is straightforwardly obtained to be
\begin{equation}\label{field_density_matrix}
    \begin{split}
        \hat \rho_F(t)
        &= |N_{\mathrm{eg}}|^{-2}[|\alpha_e|^2(\hat U_{11}|\gamma_e\rangle\langle \gamma_e|\hat U_{11}^\dagger + \hat U_{21}|\gamma_e\rangle\langle \gamma_e|\hat U_{21}^\dagger) \\
        &+\alpha_e \bar\alpha_g(\hat U_{11}|\gamma_e\rangle\langle \gamma_g|\hat U_{12}^\dagger + \hat U_{21}|\gamma_e\rangle\langle \gamma_g|\hat U_{22}^\dagger)\\
        &+\bar\alpha_e \alpha_g(\hat U_{12}|\gamma_g\rangle\langle \gamma_e|\hat U_{11}^\dagger + \hat U_{22}|\gamma_g\rangle\langle \gamma_e|\hat U_{21}^\dagger)\\
        &+|\alpha_g|^2(\hat U_{12}|\gamma_g\rangle\langle \gamma_g|\hat U_{12}^\dagger + \hat U_{22}|\gamma_g\rangle\langle \gamma_g|\hat U_{22}^\dagger)],
    \end{split}
\end{equation}
with $|\gamma_e\rangle=\sum_{n=0}^\infty c_n|n\rangle$ and $|\gamma_g\rangle=\sum_{n=0}^\infty d_n|n\rangle$. An explicit expression for the Wigner function is then directly computed as
\begin{widetext}
    \begin{equation}\label{wigner_f}
    \begin{split}
        W (\alpha,\alpha^*)= \frac{e^{-|\alpha|^2}}{|N_{\mathrm{eg}}|^2 \pi}\sum_{s=0}^\infty (-1)^s \left\{ |\alpha_e|^2\left(|r_1|^2 + |r_2|^2\right)+ |\alpha_g|^2\left(|r_3|^2 + |r_4|^2\right)
         +2\mathrm{Re}\left[\alpha_e \bar\alpha_g \left( r_1 \bar r_3 + r_2 \bar r_4 \right)\right]\right\},
    \end{split}
\end{equation}
\end{widetext}
where we have defined the functions
\begin{equation}
    \begin{split}
        & r_1(s) = \sum_{n=0}^\infty c_n \mathcal E_{n-k} \bar{\mathcal F}_{n-k} \mu(\alpha,n,s),\\
        & r_2(s) = \sum_{n=0}^\infty c_{n+k} \mathcal E_n \mathcal G_n \sqrt{ \frac{(n+k)!}{n!} } \mu(\alpha,n,s),\\
        & r_3(s) = \sum_{n=0}^\infty d_n \mathcal E_n \mathcal G_n \sqrt{ \frac{(n+k)!}{n!} } \mu(\alpha,n+k,s),\\
        & r_4(s) = \sum_{n=0}^\infty d_n \mathcal E_n \mathcal F_n \sqrt{ \frac{(n+k)!}{n!} } \mu(\alpha,n,s),
    \end{split}
\end{equation}
as well as
\begin{equation}
    \mu(\alpha,n,s) = \bar\alpha^{n-s} \sqrt{\frac{s!}{n!}} L_s^{n-s}(|\alpha|^2),
\end{equation}
with $L_x^y$ the associated Laguerre polynomials of order $x$ and $\mathcal E_n = \langle n|\mathcal E_{\hat n}|n\rangle$.
Additionally, $\mu$ fulfills the useful property $\mu(\alpha,n,s) = \bar\mu(-\alpha,s,n)$, inherited from the associated Laguerre polynomials.

\subsection{Fidelity}
Besides, the fidelity of $|\phi(t)\rangle$ with respect to the initial state $|\phi(0)\rangle$ is to be considered as well. In general, the fidelity between two states (density matrices) $\hat\rho$ and $\hat\sigma$ is defined as \cite{Nielsen_book}
\begin{equation}\label{fid}
    F(\hat\rho, \hat\sigma) = \mathrm{Tr}\sqrt{ \sqrt{\hat\rho} \hat\sigma \sqrt{\hat\rho}}.
\end{equation}
If $\hat\rho$ is a pure state ($\hat\rho^2 = \hat\rho$, $\mathrm{Tr}[ \hat\rho^2 ]= 1$), and $\hat\sigma$ is arbitrary, the fidelity (\ref{fid}) takes on the simplified form (using the notation of Nielsen and Chuang)
\begin{equation}\label{fids}
F(|\psi\rangle, \hat\sigma) = \sqrt{\langle\psi|\hat\sigma|\psi\rangle},
\end{equation}
where $\hat\rho = |\psi\rangle\langle\psi|$ is the pure state. In the case of an initial separable state $|\phi(0)\rangle = |\phi_A(0)\rangle \otimes |\phi_F(0)\rangle$, with the subindices standing for ``Atom" and ``Field", respectively, the fidelity $F(|\phi_F(0)\rangle,\hat\rho_F(t))$ between the initial and evolved cavity-field states can be straightforwardly obtained from the cavity-field reduced density matrix in (\ref{field_density_matrix}), with $|\gamma_e\rangle = |\gamma_g\rangle \equiv |\phi_F(0)\rangle$, and the expression (\ref{fids}) as
\begin{widetext}
    \begin{equation}\label{fid_sep}
        F = \frac{1}{N_{\mathrm{eg}}}\left\{ |\alpha_e|^2\left(|h_1|^2 + |h_2|^2\right)+ |\alpha_g|^2\left(|h_3|^2 + |h_4|^2\right)
         +2\mathrm{Re}\left[\alpha_e \bar\alpha_g \left( h_1 \bar h_3 + h_2 \bar h_4 \right)\right] \right\}^{1/2},
    \end{equation}
\end{widetext}
where the following definitions
\begin{equation}
    \begin{split}
        h_1 &= \sum_{n=0}^\infty |c_n|^2 \mathcal E_{n-k} \bar{\mathcal F}_{n-k},\\
        h_2 &= \sum_{n=0}^\infty \bar c_n c_{n+k} \sqrt{\frac{(n+k)!}{n!}}\mathcal G_n \mathcal E_n,\\
        h_3 &= \sum_{n=0}^\infty \bar c_{n+k} c_n \sqrt{\frac{(n+k)!}{n!}}\mathcal G_n \mathcal E_n,\\
        h_4 &= \sum_{n=0}^\infty |c_n|^2 \mathcal E_n \mathcal F_n,
    \end{split}
\end{equation}
were used. Note that, depending on the author, the fidelity is sometimes referred to as the square of the right-hand side of (\ref{fid}). However, the definition given in Ref. \cite{Nielsen_book} is employed in this article.

In the next section, we provide specific numerical examples of the results obtained in the present one. Such numerical results make evident interesting phenomena taking place in the counter-rotating system defined by the Hamiltonian (\ref{multi_kerr}), and appearing naturally as a consequence of the SUSY mapping developed in the context of the current work.

\section{Numerical results}\label{sec_4}

In this section, particular numerical examples for the dynamical variables studied in the previous one are given. The aim of this section is two-fold: first, we want to prove the applicability of the results developed up to now, and second, we want to highlight interesting features of the counter-rotating system that might have potential applications in technology development.

Note that the analytical results given in section \ref{sec_3} were tested by comparison with the numerical ones obtained with the Quantum Toolbox in Python (QuTiP \cite{Johansson2013}). For both the numerical and analytical results the base of the Fock space was truncated at $N = 350$, after having performed a convergence test. Besides, in all the simulations a weak-coupling regime is considered, i.e. $g = 0.1 \, \omega_c$ is used, the cavity-field frequency sets the time scale as $\omega_c t$ and the atomic and cavity-field frequencies are considered to be equal, this is $\Delta = 0$.

In addition, the nonlinear Kerr and the multiphotonic scenarios are analyzed separately, to evaluate better their isolate effects on the qubit and/or the cavity-field dynamics.

\subsection{Multiphotonic linear scenario}
\begin{figure}[t!]\
\centering
\includegraphics[width=1\linewidth]{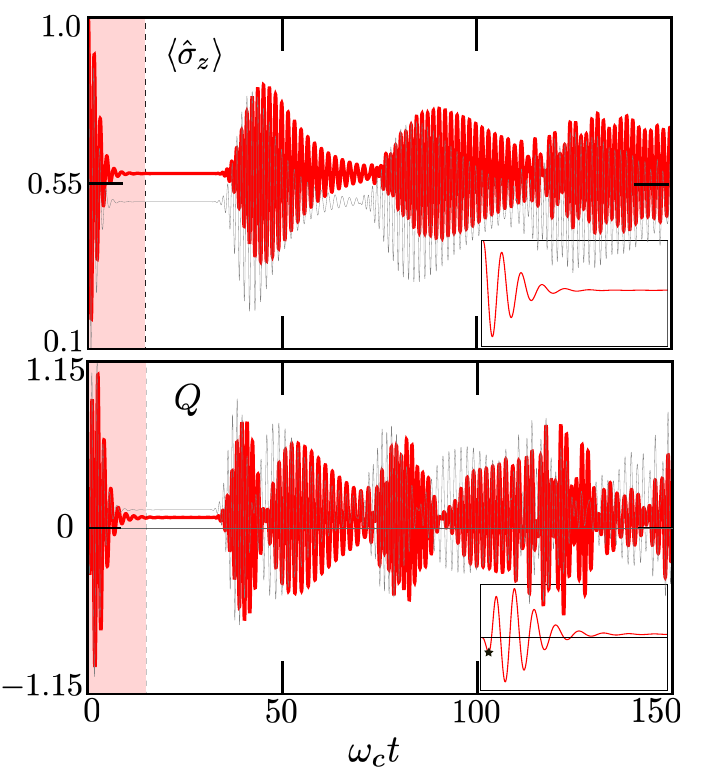}
\caption{\label{fig_2}  {\bf Atomic inversion and Mandel $Q$ parameter for $|\phi(0)\rangle = |e\rangle\otimes |\gamma\rangle$ in the multiphoton scenario.} The atomic inversion (upper panel) shows typical collapse-revival behavior. In turn, the Mandel $Q$ parameter (lower panel) indicates the non-classical (sub-Poissonian) nature of light when it takes negative values. It has been taken $g = 0.1\,\omega_c$, $\omega_a = \omega_c$, $k=2$, $\gamma = \sqrt{14}$ (red) and $\gamma = 4$ (gray). The insets show the behavior at short interaction times (i.e. the shaded area). Although $Q$ increases with $\gamma$, its first local minima (star in the inset) is shown to be negative in a wide range of the involved parameters (Figure \ref{fig_3}) for the chosen initial condition, indicating \textit{robust} non-classical nature of the electromagnetic field in the cavity produced by the interaction with the qubit at short interaction times.} 
\end{figure}

\begin{figure}[t!]\
\centering
\includegraphics[width=1\linewidth]{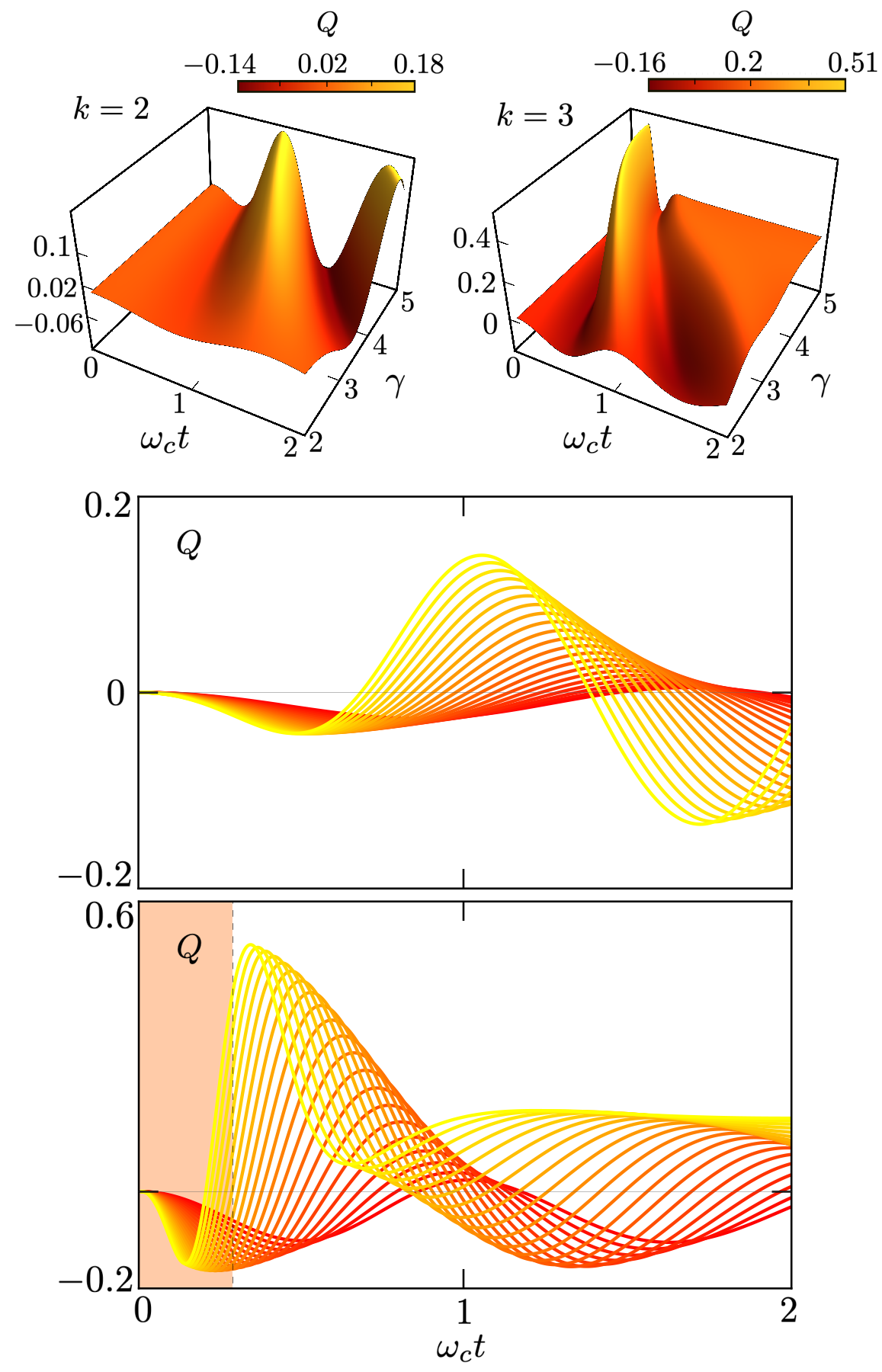}
\caption{\label{fig_3}  {\bf Mandel $Q$ parameter for $|\phi(0)\rangle = |e\rangle\otimes |\gamma\rangle$ in the multiphoton scenario for short interaction times.} Mandel $Q$ parameter as function of time $\omega_c t$ and the initial coherent field-amplitude $\gamma$ (upper panel), for $k = 2$ (upper-left) and $k = 3$ (upper-right), in the resonant case ($\omega_a = \omega_c$), $g = 0.1\,\omega_c$, for short interaction times. Transversal cuts (middle and lower panels, corresponding to $k=2$ and $k=3$, respectively) for discrete values of $\gamma$, ranging from $\gamma = 2.0$ (red) to $\gamma = 4.0$ (yellow) in steps of $0.1$. The first local minimum is seen to \textit{always} take on negative values. Similar behavior is obtained for the non-resonant situation $\omega_a\neq\omega_c$, and different values of the remaining parameters.} 
\end{figure}

\begin{figure}[t!]\
\centering
\includegraphics[width=1\linewidth]{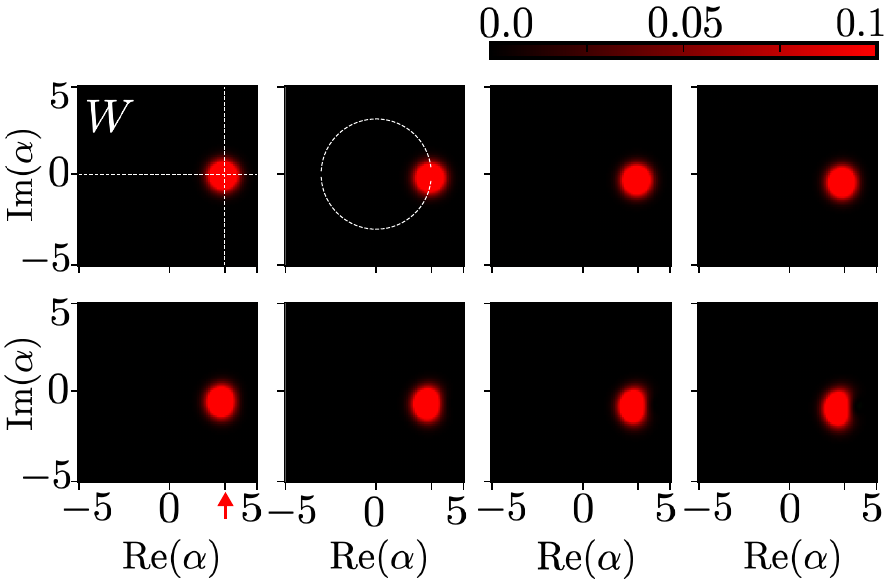}
\caption{\label{fig_4}  {\bf Wigner function $W$ dynamics for $|\phi(0)\rangle = |e\rangle\otimes |\gamma\rangle$ in the multiphoton scenario for short interaction times.} From left to right, first the upper row and after the lower one, it is shown the evolution of the Wigner function (in phase space) of the cavity-field reduced density matrix, from $t = 0$ to $t = 0.28\,\omega_c^{-1}$ in steps of $0.04\,\omega_c^{-1}$, for the resonant case $k = 3$, $\gamma = 3.1$, $g = 0.1\,\omega_c$. At $t=0$ (first panel) the Wigner function is a perfect ball centered at $\alpha = \gamma$ (the red arrow indicating this value). As $t$ increases the Wigner function moves clockwise along a circle of radius $\gamma$ centered at the origin of the complex plane (phase space). This is represented by the white circle in the second panel. At $t=0.28\,\omega_c^{-1}$ (last panel, lower row) the Wigner function shows clear squeezing effects due to the interaction with the atom. This is in agreement with the negative values of the Mandel $Q$ parameter at $t=0.28\,\omega_c^{-1}$ (dashed vertical line delimiting the shaded area in the lower panel of Figure \ref{fig_3}). } 
\end{figure}

Figure \ref{fig_2} shows the atomic inversion $\langle\hat\sigma_z\rangle$ (upper panel) and the Mandel $Q$ parameter (lower panel) as functions of the dimensionless time $\omega_c t$, according to (\ref{expec_diagonal}) and Table \ref{table1}, in the case of the linear ($\chi = 0$) multiphotonic ($k = 2$) counter-rotating system, for the initial condition $|\phi(0)\rangle = |e\rangle \otimes |\gamma\rangle$, where $|\gamma\rangle$ is a coherent state and specific values of the remaining parameters. This is $\alpha_g = 0$ was taken, together with $c_n = \exp(-|\gamma|^2/2)\gamma^n/n!$. Both $\langle\hat\sigma_z\rangle$ and $Q$ show typical collapse-revival behavior, as expected; $\langle \hat n \rangle$ is not shown in Figure \ref{fig_2}, however, it has the same form shown by $\langle\hat\sigma_z\rangle$ due precisely to $\hat C$. Indeed, $Q$ shows local minima taking on negative values, and therefore reflecting the sub-Poissonian (quantum) nature of the cavity field (see the inset in the lower panel of Figure \ref{fig_2}). In particular, the first local minimum (marked by a star) remains negative in a wide range of values of the involved parameters. This can be appreciated in Figure \ref{fig_3} for the cases $k=2$ and $k=3$. Such robustness is precisely due to the chosen initial condition and the multiphotonic ($k>1$) interaction. Moreover, this interesting behavior is also present in the case where there exists detuning between the field and atomic frequencies, i.e. $\Delta\neq 0$, as well as for different values of the additional parameters. Furthermore, the evolution of the Wigner function, according to the analytical expression provided in (\ref{wigner_f}), is presented in Figure \ref{fig_4}, for the case $k=3$ and short interaction times (the time interval corresponding to the shaded area in the lower panel of Figure \ref{fig_3}). At $t = 0$, the Wigner function is a perfect ball centered at $\gamma = 3.1$. As $t$ increases, the ball moves clockwise along a circle of radius $\gamma$ around the origin of the phase space. Additionally, and as can be particularly appreciated in the last (lower-right) panel of Figure \ref{fig_4}, the Wigner function shows visible squeezing due to the multiphotonic ($k>1$) interaction. This is certainly in agreement with the negative values of the Mandel $Q$ parameter at short interaction times presented in the lower panel of Figure \ref{fig_3}.  In turn, squeezed states of light have a broad range of important applications; probably the most outstanding one in metrology, for the sensing of gravitational waves (see for example Refs. \cite{Ligo_2011,Tse_2019}). 

\subsection{Nonlinear Kerr medium}
In the case of the single-photon ($k = 1$) nonlinear Kerr medium ($\chi\neq 0$), both the atomic inversion $\langle \hat\sigma_z \rangle$ and the Mandel $Q$ parameter present collapse-revival behavior, as in the linear multiphoton scenario. Nonetheless, such collapse-revival behavior is the one characteristic of a nonlinear interaction: periodic complete (and perfectly distinguishable) revivals (see Refs. \cite{Buck_81,Buzek_90,Moya_95Kerr,Cordero_2011}). Such periodic complete revivals are indeed well-studied in the context of the rotating interaction, however, their appearance in the context of the counter-rotating systems is, up to the authors' knowledge, not reported in the currently available literature. It is then worthwhile to remark that perfect revivals naturally arise also in the counter-rotating interaction. In turn, this brings up the question of whether the field reproduces its initial condition periodically in time as well. The answer can be certainly obtained from the fidelity calculated in the previous section.

\begin{figure}[t!]
\centering
\includegraphics[width=1\linewidth]{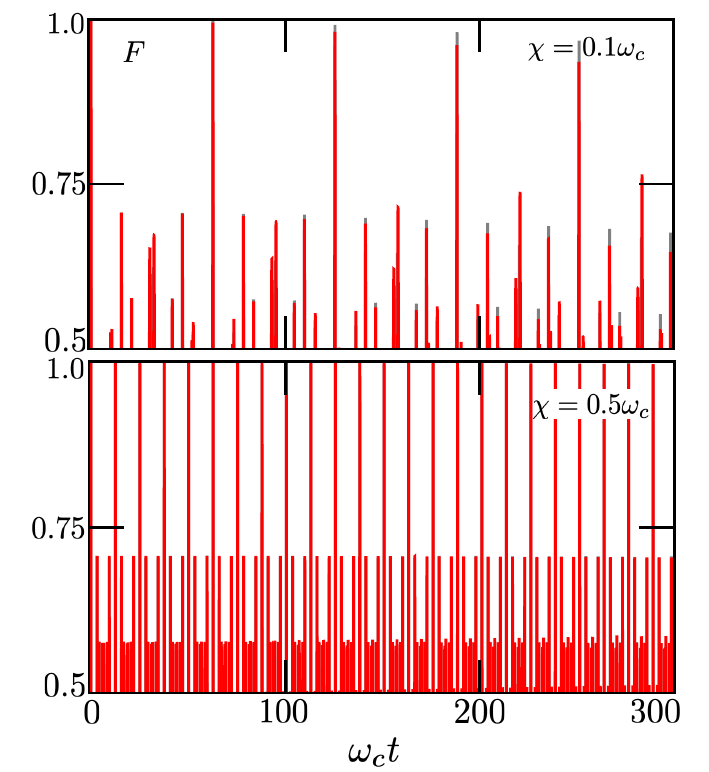}
\caption{\label{fidelity_kerr}  {\bf Fidelity $F$ dynamics for $|\phi(0)\rangle = |e\rangle\otimes |\gamma\rangle$ in the nonlinear Kerr scenario.} Fidelity $F(t) = \sqrt{ \langle\phi_F(0)|\hat\rho_F(t)|\phi_F(0)\rangle }$ as function of time in the resonant case with $g = 0.1\,\omega_c$, $\chi = 0.1\,\omega_c$ (up) and $\chi = 0.5\,\omega_c$ (down), $\gamma = 3.1$ (red) and $\gamma = 4.1$ (gray). It can be observed that for sufficiently large values of the nonlinear parameter $\chi$ (lower panel), the initial state is recovered perfectly, i.e. $F\approx 1$, in a periodic manner even at relatively large times. Besides, as the nonlinear parameter $\chi$ increases the amplitude of the coherent cavity field $\gamma$ becomes less important. } 
\end{figure}

\begin{figure}[t!]
\centering
\includegraphics[width=1\linewidth]{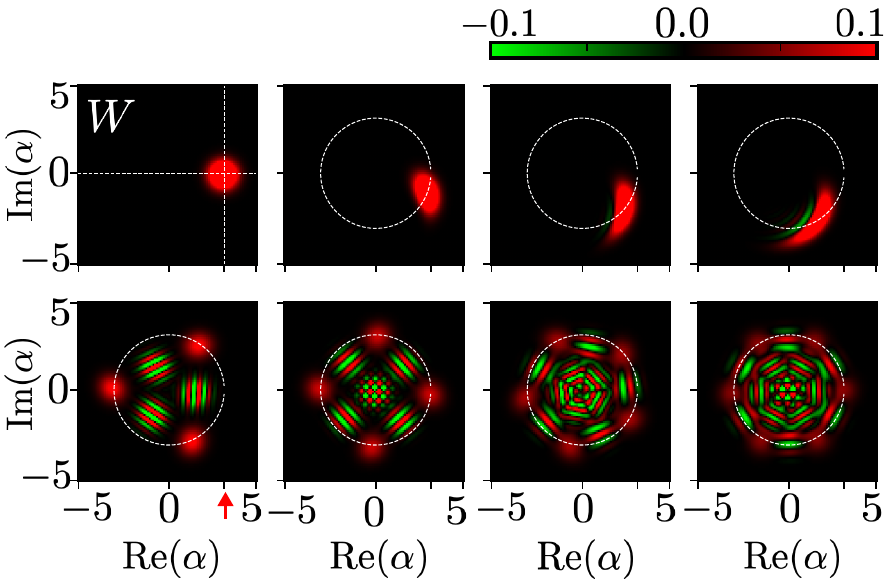}
\caption{\label{wigner_kerr}  {\bf Wigner function $W$ dynamics for $|\phi(0)\rangle = |e\rangle\otimes |\gamma\rangle$ in the Kerr scenario.} The resonant case with $g = 0.1\,\omega_c$, $\gamma = 3.1$ and $\chi = 0.5\,\omega_c$ is considered. The upper row shows the dynamics of the Wigner function at short interaction times: left to right, from $t = 0$ to $t = 0.09\,\omega_c^{-1}$ in steps of $0.03\,\omega_c^{-1}$. The typical stretching of the distribution, as it moves clockwise around a circle of radius $\gamma$, characteristic of a Kerr medium can be appreciated. The lower row shows the Wigner function for specific instants for time, from left to right: $t = 2.1\,\omega_c^{-1},11\,\omega_c^{-1},11.3\,\omega_c^{-1}$ and $17.8\,\omega_c^{-1}$. Then, at specific instants of time, generalized Schr\"odinger-cats appear, i.e. a number $\ell\in\mathbb Z$, ($\ell = 3, 4, 5, 6$ Schr\"odinger-cats are shown) of perfect balls distributed along the circumference of radius $\gamma$, together with the corresponding interference terms. The \textit{typical} Schr\"odinger cat $\ell = 2$ (not shown) also appear at particular instants of time (see \cite{Ayyash_2024}). Naturally, whenever $F = 1$ in the lower panel of Figure \ref{fidelity_kerr}, the Wigner function becomes that of the first panel in the upper row.} 
\end{figure}

For the initial pure state $|\phi(0)\rangle = |e\rangle \otimes  |\gamma\rangle$, the fidelity (\ref{fid_sep}) takes the short form $F = \sqrt{|h_1|^2 + |h_2|^2}$. Figure \ref{fidelity_kerr} shows the fidelity between the evolved and initial field states, $\hat\rho_F(t)$ and $|\phi_F(0)\rangle$, respectively, as a function of the dimensionless time $\omega_c t$, for a pair of values of the nonlinear parameter $\chi$, as well as a pair of values for the amplitude of the initial coherent field $\gamma$, for the resonant ($\Delta = 0$) single-photon scenario. It can be appreciated that, for sufficiently large $\chi$ (lower panel in Figure \ref{fidelity_kerr}), the cavity field indeed reproduces its initial condition (i.e. $F = 1$) periodically in time, even for relatively large time intervals. Moreover, the coherent field amplitude $\gamma$ seems to have less influence as the nonlinear parameter $\chi$ increases. It is worth mentioning that such behavior is present as well in the non-resonant case ($\Delta \neq 0$) and for some other values of the remaining parameters. Furthermore, the complete recovery of the initial cavity-field condition appears in the case of an initial non-separable state too, namely where both $\alpha_e$ and $\alpha_g$ are different from zero and $c_n\neq d_n$ for some $n$. In addition, the Wigner function also recovers its initial form (the perfect ball centered at $\gamma$ shown in the upper-left panel of Figure \ref{wigner_kerr}) whenever $F = 1$ in Figure \ref{fidelity_kerr}. Certainly, the evolution of the Wigner function at short interaction times is shown in the upper row of Figure \ref{wigner_kerr}. The characteristic stretching of the Wigner function consequence of a nonlinear interaction can be appreciated (compare with \cite{Werner_91,Kitagawa_86}). In addition, at specific instants of time, \textit{generalized} Schr\"odinger cats (SC), i.e. a number $\ell = 2, 3, ...$ of perfect balls distributed along a circumference of radius $\gamma$, with the corresponding interference fringes between them, appear. These are shown in the lower row of Figure \ref{wigner_kerr} (compare also with \cite{Rivera_97}). These generalized SC might be important in areas such as quantum information, see for instance \cite{Gilchrist_2004,Schlegel_2022,Ayyash_2024}, and have not been reported before in the context of the counter-rotating interaction defined by Hamiltonian (\ref{multi_kerr}).

\section{Conclusions}\label{sec_5}
A quite general interaction in the framework of a counter-rotating system has been studied. Exact analytical expressions for some of the dynamical variables of interest of the system have been provided. Such are straightforwardly obtained using the rather simple supersymmetric transformation and by departing from the known solutions of the corresponding rotating system. Two cases of interest have been considered, namely, the multiphotonic and the nonlinear Kerr scenarios. Moreover, some intriguing features of both systems have arisen naturally from the present analysis, i.e. the generation of squeezed states of light in the multiphotonic case and the generation of generalized Schr\"odinger cats in the nonlinear Kerr scheme. The former has an ample range of applications in metrology and the latter in the management of quantum information. In turn, both are important areas of applied physics.

Besides, the counter-rotating interaction investigated in this work is susceptible to be implemented in laboratory schemes by employing nowadays available technology. In turn, this opens the possibility to study counter-rotating interactions in more general scenarios. Indeed, the current study is far from exhausting the whole of possibilities. Additionally, the presented results might be also useful in modeling open systems, such as those being provided with photons from the exterior or those in which sinks of photons are accounted for, as mentioned in Reference \cite{Solano_PRL03}. Then the non-conservative (counter-rotating) terms can be interpreted as appearing due to the external source or sink of photons.

Also, it is really important to remark that the implementation of supersymmetry is almost unexplored in the contexts of cavity quantum electrodynamics and quantum optics. Nevertheless, as it has been demonstrated in this discussion, it is an advantageous tool for the link, construction, and analysis of exactly solvable models, as well as for the corresponding test of the physical repercussions and the evaluation of the feasible technological applications, and is therefore worth of attention. Finally, it would be interesting to assess the possibility of implementing the SUSY formalism in the framework of, for instance, the Lindblad master equation, or for some other known models, such as the Landau-Zener or the Dicke models.

\section*{Acknowledgments}
The work of I. A. B.-G. is supported by Spanish MCIN with funding from European Union Next Generation EU (PRTRC17.I1) and Consejeria de Educacion from Junta de Castilla y Leon through QCAYLE project, as well as Grant No. PID2023-148409NB-I00 MTM funded by AEI/10.13039/501100011033, and RED2022-134301-T. Financial support of the Department of Education, Junta de Castilla y Le\'on, and FEDER Funds is also gratefully acknowledged (Reference: CLU-2023-1-05). In addition, L. Hernández-Sánchez acknowledges the Instituto Nacional de Astrofísica, Óptica y Electrónica (INAOE) for the collaboration scholarship granted and the Consejo Nacional de Humanidades, Ciencias y Tecnologías (CONAHCYT) for the SNI Level III assistantship (CVU No. 736710).

\appendix

\section{Matrix representation: a useful one}\label{appendix_matrix}
In this Appendix, a representation in the qubit base $\left\{|e\rangle,|g\rangle\right\}$ of the operators and states in the main text is given. The Hamiltonian (\ref{gen_H}) can be cast in a matrix form as
\begin{equation}
    \hat H_0 
    = \begin{pmatrix}
    \frac{\Delta}{2} + D_{\hat n}^+ & & g\hat a^k f_{\hat n}\\
    gf_{\hat n}(\hat a^\dagger)^k & & - \frac{\Delta}{2} + D_{\hat n}^- 
    \end{pmatrix},
\end{equation}
where we have defined $D_{\hat n}^\pm \equiv G_{\hat n} \pm F_{\hat n}$. The SUSY operator can be expressed in a simple form
\begin{equation}
    \mathcal{\hat B}^k
    = \begin{pmatrix}
    \hat a^{\dagger k} & 0\\
    0 & \hat {a}^k
    \end{pmatrix},
\end{equation}
and by considering (\ref{susy_t}), the transformed Hamiltonian $\hat H$ is straightforwardly seen to be
\begin{equation}
\hat H 
= \begin{pmatrix}
\frac{\Delta}{2} + D_{\hat n - k\mathbb I}^+ & &gf_{\hat n}(\hat a^\dagger)^k\\
g\hat a^k f_{\hat n} & &- \frac{\Delta}{2} + D_{\hat n+k\mathbb I}^-
\end{pmatrix},
\end{equation}
where we have used the fact that $\hat a^k g_{\hat n} = g_{\hat n +k\mathbb I}\hat a^k$, with $g_{\hat n}$ an arbitrary function of the number operator $\hat n$. Additionally, in section \ref{sec_3} the initial normalized condition
\begin{equation}
    |\phi(0)\rangle
    = \frac{1}{N_{\mathrm{eg}}} \begin{pmatrix}
    \alpha_e \sum_n c_n(0)|n\rangle\\
    \alpha_g \sum_n d_n(0)|n\rangle
    \end{pmatrix},
\end{equation}
is considered for the counter-rotating system. For consistency (we dig further into this point in Appendix \ref{appendix_entanglement}), the corresponding non-normalized initial condition of the rotating system shall be
\begin{equation}
    |\psi(0)\rangle
    = \frac{1}{N_{\mathrm{eg}}} \begin{pmatrix}
    \alpha_e \hat a^k \frac{(\hat n - k \mathbb I)!}{\hat n!}\sum_n c_n(0)|n\rangle\\
   \alpha_g (\hat a^\dagger)^k \frac{\hat n!}{(\hat n + k \mathbb I)!}\sum_n d_n(0)|n\rangle
    \end{pmatrix}.
\end{equation} 
This simple representation allows us to perform the calculations simply and transparently and is doubtlessly valuable.

\section{Propagators for the multiphotonic nonlinear Kerr JC model}\label{appendix_solutions}
The propagator $\mathcal{\hat U}(t) = \exp(-i\hat H_0 t)$, giving the temporal evolution associated with the Hamiltonian
\begin{equation}\label{Kerrdisp}
    \hat H_0 = \frac{\Delta}{2}\hat\sigma_z + \hat Z_+ + \hat Z_- + g[\hat a^k \hat\sigma_+ +  (\hat a^\dagger)^k\hat\sigma_-],
\end{equation}
where we have defined $\hat Z_\pm \equiv \hat\sigma_\pm \hat\sigma_\mp G_{\hat n \pm k\mathbb I}$, with $G_{\hat n} = \chi \hat n^2$, can be obtained (up to a global phase) as
\begin{equation}\label{evopH}
\begin{split}
    & \mathcal{\hat{U}} = \left(
\begin{matrix} 
\mathcal{\hat U}_{11} & \mathcal{\hat U}_{12}
\\ 
\mathcal{\hat U}_{21} & \mathcal{\hat U}_{22}
\end{matrix}
\right),\\
\mathcal{\hat U}_{11} =  \mathcal E_{\hat n} &\mathcal F_{\hat n}^\dagger, \qquad\mathcal{\hat U}_{22} = \mathcal E_{\hat n- k\mathbb I} \mathcal F_{\hat n- k\mathbb I},\\
\mathcal{\hat U}_{12} = \hat a^k \mathcal E_{\hat n - k\mathbb I} &\mathcal G_{\hat n - k\mathbb I},\qquad\mathcal{\hat U}_{21} = (\hat a^\dagger)^k\mathcal E_{\hat n} \mathcal G_{\hat n},
\end{split}
\end{equation}
where
\begin{equation}
    \begin{split}
        \mathcal F_{\hat n} & = \cos(\Omega_{\hat n}t) + i \left[\frac{\Delta}{2} + \chi(\hat nk + \frac{k^2}{2}\mathbb I)\right]\frac{\sin(\Omega_{\hat n}t)}{\Omega_{\hat n}},\\
        \mathcal E_{\hat n} & =  \exp\left[-i\chi t\hat n(\hat n + k\mathbb I)\right],\qquad \mathcal G_{\hat n} = -ig \frac{\sin(\Omega_{\hat n})t}{\Omega_{\hat n}}, \\
        \Omega_{\hat n}^2 & = g^2\frac{(\hat n + k\mathbb I)!}{\hat n!} +\left[\frac{\Delta}{2} + \chi(\hat n k + \frac{k^2}{2}\mathbb I) \right]^2.
    \end{split}
\end{equation}
Note that the modified Rabi frequency operator $\Omega_{\hat n}$ depends on both $k$ and the nonlinear parameter $\chi$, however to keep the notation as simple as possible we only make explicit its dependence on the number operator $\hat n$.

It is important to remark that in this Appendix the evolution operator associated with (\ref{Kerrdisp}) is presented due to the displacements generated in the functions $F_{\hat n}$ and $G_{\hat n}$ by the application of the intertwining relation (\ref{susy_t}), as can be appreciated in (\ref{SUSYp}) through the $\hat S_\pm$ operators. However, such can be compensated in the initial Hamiltonian $\hat H_0$, as it was done in (\ref{Kerrdisp}), to produce the desired counter-rotating Hamiltonian given in (\ref{multi_kerr}).\\

\section{(Non-)Separability from the SUSY operator}\label{appendix_entanglement}
By virtue of the intertwining relation (\ref{susy_t}), the solution $|\phi(t)\rangle$ of the Schr\"odinger equation associated with $\hat H$ can be expressed in two different but equivalent manners
\begin{equation}\label{phi_t}
    |\phi(t)\rangle = \left\{
    \begin{matrix}
        \mathcal{\hat A}|\psi(t)\rangle \\
        \exp(-i\hat H t) \mathcal{\hat A} |\psi(0)\rangle
    \end{matrix}
    \right. .
\end{equation}
The state $|\psi(t)\rangle$ is assumed to be non-normalized. The first expression in (\ref{phi_t}) is the one below (\ref{susy_t}) in the main text. However, as $\mathcal{\hat A} |\psi(0)\rangle = |\phi(0)\rangle$, the second expression in (\ref{phi_t}) is the one used in section \ref{sec_3} and the remaining of the main text, and reveals a remarkable property of the SUSY connection between $\hat H_0$ and $\hat H$.

The initial condition $|\psi(0)\rangle$ associated with the rotating system $\hat H_0$ is set to be 
\begin{equation}\label{initial_rotating}
    \begin{split}
        |\psi(0)\rangle = N_{\mathrm{eg}}^{-1}[\alpha_e|e\rangle\otimes\hat a^k\frac{(\hat n-k\mathbb I)!}{\hat n!}\sum_{n=0}^\infty c_n(0)|n\rangle \\
        +\; \alpha_g|g\rangle\otimes(\hat a^\dagger)^k\frac{\hat n!}{(\hat n+k\mathbb I)!}\sum_{n=0}^\infty d_n(0)|n\rangle],
    \end{split}
\end{equation} 
with $N_{\mathrm{eg}}=\sqrt{ |\alpha_e|^2 + |\alpha_g|^2}$, $\sum_{n=0}^\infty |c_n(0)|^2 = \sum_{n=0}^\infty |d_n(0)|^2 =1$, and such that $|\phi(0)\rangle = \mathcal{\hat B}^k|\psi(0)\rangle$ is normalized and simply given by
\begin{equation}\label{initial_counter}
    \begin{split}
        |\phi(0)\rangle = N_{\mathrm{eg}}^{-1}[\alpha_e|e\rangle&\otimes\sum_{n=0}^\infty c_n(0)|n\rangle\\
    + \; \alpha_g|g\rangle&\otimes\sum_{n=0}^\infty d_n(0)|n\rangle].
    \end{split}
\end{equation}
The relations $(\hat a^\dagger)^k \hat a^k = \frac{\hat n!}{(\hat n - k \mathbb{I})!}$ 
and $\hat a^k (\hat a^\dagger)^k= \frac{(\hat n + k \mathbb{I})!}{\hat n!}$ were used. The initial state (\ref{initial_counter}) is a non-separable state of the qubit-cavity counter-rotating system as long as $c_n(0)\neq d_n(0)$, for any $n$. However, if $c_n(0)= d_n(0)$ for all $n$, the state (\ref{initial_counter}) becomes separable even when $|\psi(0)\rangle$ in (\ref{initial_rotating}) is not. In that sense, the intertwining operator $\mathcal{\hat B}^k$ produces the separability of the state (\ref{initial_counter}). It is also possible to choose a separable $|\psi(0)\rangle$ and such that $|\phi(0)\rangle$ is non-separable anymore. Therefore, the intertwining operator changes the \textit{nature} (separable or not) of the state it acts on. \\

In particular, the choice of $|\psi(0)\rangle$ in (\ref{initial_rotating}), allows us to consider a general non-separable state $|\phi(0)\rangle$ as the initial state for the counter-rotating system, that is the object of analysis of this manuscript. As mentioned, for the case $c_n(0)= d_n(0)$ for all $n$, the state (\ref{initial_counter}) becomes separable and normalized, which might be a more suitable description if the qubit and the cavity are not supposed to be entangled at $t=0$, which in turn could be a better picture in some specific scenarios.

\bibliography{My_Bib_SUSY_JC}
\end{document}